\documentclass[12pt]{iopart}
\usepackage{iopams,epsfig}
\begin{document}
\title{Parity Breaking in Nematic Tactoids}

\author{P Prinsen\dag\ and P van der Schoot\ddag}
\address{\dag\ Complex Fluids Theory, Faculty of Applied Sciences,
Delft University of Technology, P.O. Box 5057, 2600 GB Delft, The
Netherlands}
\address{\ddag\ Eindhoven Polymer Laboratories,
Eindhoven University of Technology, P.O. Box 513, 5600 MB
Eindhoven, The Netherlands}

\ead{p.prinsen@tnw.tudelft.nl}

\begin{abstract}
We theoretically investigate under what conditions the director
field in a spindle-shaped nematic droplet or tactoid obtains a
twisted, parity-broken structure. By minimizing the sum of the
bulk elastic and surface energies, we show that a twisted director
field is stable if the twist and bend elastic constants are small
enough compared to the splay elastic constant, but only if the
droplet volume is larger than some minimum value. We furthermore
show that the transition from an untwisted to a twisted
director-field structure is a sharp function of the various
control parameters. We predict that suspensions of rigid, rod-like
particles cannot support droplets with a parity broken structure,
whereas they could possibly occur in those of semi-flexible,
worm-like particles.
\end{abstract}

\submitto{JPCM}


\maketitle

\section{Introduction}

Tactoids are metastable nematic liquid crystalline droplets
suspended in the co-existing isotropic fluid parent phase
\cite{Bern1,Zocher,Bern2,Dogic,Drzaic1,Drzaic2,Sonin}. They can
adopt a variety of shapes and director-field configurations,
depending on their size, on their elastic properties and on the
degree of anisotropy of the surface tension
\cite{Prinsen1,Prinsen2,Williams1,Williams2,Bogdanov,Kaznacheev,Virga,Chandr,Bates,Care1,Care2}.
Theoretically, small nematic droplets tend to be quite elongated
with a director field that is uniform, whereas very large droplets
should be more or less spheroidal with a bipolar director field
that supports two surface defects called boojums \cite{Prinsen1}.
The transformation of the shape and the director-field
configuration of the droplets with increasing size appears to be
continuous \cite{Prinsen2}. It should be noted that tactoids are
quite different from the usual polymer dispersed liquid crystals
or PDLCs, in which nematic droplets are embedded in a (solid)
polymeric matrix. Indeed, in PDLCs the liquid-crystalline droplets
have little opportunity to adjust their shape once the polymer
matrix has set. Their surface anchoring is usually also fixed by a
strongly anisotropic surface tension that is not nearly as small
as seems to be the case in tactoids \cite{Prinsen1,Amb}.

In 1985, Williams showed theoretically that spherical bipolar
droplets can also exhibit a parity-broken, twisted director field
without the nematogens needing to be chiral \cite{Williams1}.
Twist transitions have in fact also been predicted to occur in
other geometries including nematic fluids confined between two
concentric spheres \cite{Stark} and in so-called hybrid nematic
layers \cite{Perg,Perg1,Kis}. In a twisted nematic droplet the
director field circles around the main droplet axis as shown in
Figure 1. Twisted bipolar droplets have indeed been observed in
thermotropic liquid crystals \cite{Drzaic1,Volovik,Kurik}. To our
knowledge, they have not been observed in lyotropic liquid
crystals---as we shall see, this may be connected with their
typical elongated, spindle-like shape.

In this paper, we extend the work of Williams by considering
tactoids of finite size, that is, we allow for nematic droplet
shapes that are not necessarily spherical, and for director fields
that are not truly bipolar. Our aim is to estimate the minimum
size for the twisted director field to be stable in relation to
the elastic properties of the nematic and the interfacial tension
between the co-existing nematic and isotropic phases. To this end, we
first, in section 2, write down a phenomenological free energy
consisting of the sum of the usual Frank energy of the elastic
deformation of the director field, and a surface free energy of
the Rapini-Papoular type \cite{Rapini}, and discuss the earlier
work of Williams.

Next, in section 3, we present our parametrization of the problem,
which is based on one that we introduced in previous work, and
that allows for (but does not impose) a continuous transformation
from a bipolar to a uniform director field, and from a spherical
to a highly elongated spindle-like droplet. Here, we add an extra
degree of freedom to describe a potentially twisted director
field, being the maximum surface twist angle, as parameter. In
section 4, we first discuss truly bipolar droplets, which can
largely be dealt with analytically, and generalize Williams'
result to bipolar droplets of arbitrary aspect ratio. As we shall
see, the larger their aspect ratio, the less likely a twisted
director-field configuration. We numerically investigate the more
general case of droplets that can optimize their structure within
a prescribed class of droplet shape and director-field geometry,
in section 5. We find that the transition to a twisted structure,
if it occurs, is sharp at a critical dimensionless volume $v_{0}$
and that the droplet is twisted for all volumes $v>v_{0}$ greater
than that. The critical volume $v_{0}$ is a complex function of
the various elastic constants and of the anisotropic surface
tension. We find that for lyotropic liquid crystals the twist
transition must occur at droplet volumes three to six orders of
magnitude larger than that for thermotropic liquid crystals. This
presumably explains why twisted structures have not yet been
observed in the former, whereas they have in the latter. Finally,
in the last section, section 6, we summarize our conclusions.

\section{Free Energy}

Presuming the bulk free energy describing the thermodynamic
stability of a nematic droplet to be independent of its
macroscopic structure, the excess free energy $F$ relevant to the
problem in hand may be written as \cite{Virga}
\begin{eqnarray}
F=F_{E}+F_{S} \label{frank},\\\fl F_{E}=\frac{1}{2}\int
_{V}d^{3}{\mathbf r}\left[K_{11}\left( \nabla\cdot{\mathbf
n}\right)^{2} + K_{22}\left( {\mathbf n}\cdot \nabla\times{\mathbf
n}\right)^{2} + K_{33}\left( {\mathbf n}\times
(\nabla\times{\mathbf n})\right)^{2}\right], \label{blk}
\\ F_{S}=\tau\int_{S}d^{2}{\mathbf r}\left(1+\omega ({\mathbf
q}\cdot{\mathbf n})^{2}\right),
\end{eqnarray}
where the volume integral $F_{E}$ accounts for a possible elastic
deformation of the director field, and the surface integral
$F_{S}$ for the interfacial free energy. Here, ${\mathbf
n}({\mathbf{r}})$ denotes the director field at position
${\mathbf{r}}$, ${\mathbf q}={\mathbf q}({\mathbf{r}})$ is the normal to the
droplet surface, $K_{11}$, $K_{22}$ and $K_{33}$ are the familiar
Frank elastic constants for splay, twist and bend deformations,
$\tau$ is the isotropic interfacial tension and $\omega\geq 0$ a
dimensionless anchoring strength, penalizing a nonplanar alignment
of the director field to the interface.

In equation (\ref{blk}), we ignore the contributions of the
surface elastic constants $K_{13}$ and $K_{24}$, associated with
splay-bend and saddle-splay deformations. The splay-bend
contribution we drop for simplicity, because keeping this term
would, in principle, require the incorporation of higher order
gradients in the director field to keep the problem well posed
\cite{Polak}. We note in passing that Yokoyama recently argued on
quite general grounds that $K_{13}$ should in fact vanish
identically \cite{Yokoyama}. As for the saddle-splay contribution,
it merely renormalizes the splay elastic constant $K_{11}$
\cite{Prinsen2}, at least for twist-free, bispherical
director-field configurations, in which case $K_{11}\rightarrow
K_{11}-K_{24}$ \cite{Prinsen2}. In a twisted bipolar configuration
this is no longer strictly accurate but we do not expect that
accounting for the splay-bend elasticity would qualitatively
affect the stability of the twisted director field for reasons
that we discuss in more detail in section 4. It should be stressed
that in certain geometries, such as in hybrid nematic cells with a
homeotropic alignment on one boundary surface and planar alignment
on the other, or in nematic layers with a mutually perpendicular
planar alignment on both surfaces, the saddle splay contribution
can induce a twist transition \cite{Perg,Perg1,Kis}, so the issue
does indeed deserve some attention.

The optimal droplet shape and director-field configuration
minimizes equation (\ref{frank}) at constant volume $V$ of the
drop and at constant unit length of the director
$|\mathbf{n}\mathrm{(}\mathbf{r}\mathrm{)|^{2}=1}$. This highly
non-trivial free boundary problem can be simplified considerably
in the limit of $V/\tau K_{ii}\omega\rightarrow\infty$, because
then the droplet is spherical and the director field is aligned
tangentially to the interface \cite{Prinsen2}. Tacitly presuming
this limit to hold, Williams showed that for the twisted bipolar
configuration to be favored, the elastic constants of the nematic
would have to obey the inequality \cite{Williams1}
\begin{equation}
\gamma_{33}\leq
\frac{4\pi^{2}-16}{20-\pi^{2}}(1-\gamma_{22})\approx
2.32(1-\gamma_{22}), \label{wil}
\end{equation}
where $\gamma_{22}\equiv K_{22}/K_{11}$ and $\gamma_{33 }\equiv
K_{33}/K_{11}$. This inequality expresses the circumstance that by
adopting a twisted configuration, a droplet can lower its splay
energy albeit at the cost of raising its bend and twist energy.
Hence, if the bend and twist elastic constants are small enough
compared to the splay elastic constant, a twisted configuration
can become energetically favorable. Volovik and Lavrentovich were
the first to observe a twisted bipolar droplet in a thermotropic
nematic liquid crystal \cite{Volovik}. Available experimental data
seem to indicate that equation (\ref{wil}) indeed holds
\cite{Drzaic2}.

Whilst twisted bipolar drops have been observed in thermotropic
nematics, such is as far as we are aware not the case for
lyotropic ones. This may, of course, be linked to them not obeying
equation (\ref{wil}). Indeed, theoretical calculations show that
for hard, rod-like particles, $\gamma_{22}\approx 1/3$ and
$\gamma_{33}\gtrapprox 5$ at least if they are sufficiently
slender \cite{Vroege,Lee2,Odijk}. These predictions agree with
experimental data on aqueous dispersions of tobacco mosaic virus
particles that to a first approximation behave like hard rods,
with $\gamma_{22}\approx 0.2-0.4$ and $\gamma_{33}\approx 9-17$
\cite{Hurd,Fraden}. So, it seems that a nematic of rod-like
particles interacting via purely repulsive interactions do not
obey the Williams inequality, equation (\ref{wil}), and hence
should not display a twist transition. On the other hand, one
would expect nematic droplets in dispersions of semi-flexible
particles to be able to support a twisted director field, because
calculations show that $\gamma_{22}\approx 1/3$ and
$\gamma_{33}\approx 1$ \cite{Vroege,Sato,Odijk}. This is in
qualitative agreement with experimental data albeit that the ratio
of the twist and splay constant is in practice somewhat lower than
predicted, $\gamma_{22}\approx 0.04-0.15$
\cite{Lee,Sato,Itou,Taratuta,Taratuta2,DuPre}.

The question thus arises why twisted textures have not been
observed in lyotropic nematics. To answer this question, we
generalize equation (\ref{wil}) to nonspherical, spindle-like
droplets, and allow for director-field configurations that are not
necessarily truly bipolar. As shall become clear, both finite-size
effects and non-sphericity (typical of lyotropic nematic tactoids)
conspire against the twisted state.

\section{Parametrization}

Following the prescription of references
\cite{Prinsen1,Prinsen2,Williams1}, we impose the geometry of the
director field and that of the droplet shape. See figure 2. They
are a function of three free parameters, being the aspect ratio of
the droplet, the position of the foci of the bispherical director
field \cite{Prinsen1,Prinsen2,Bijnen} and the surface twist angle
\cite{Williams1}. The director field is chosen in such a way that
it can transform continuously from a homogeneous to a bipolar
field and from an untwisted to a twisted one. The droplet shape
varies continuously from spherical to spindle like, i.e.,
elongated with sharp ends. For the given director fields and
droplet shapes, discussed in more detail below, we calculate the
total free energy, equation (\ref{frank}), and minimize it with
respect to the free parameters.

Inspired by the works of Williams \cite{Williams1} and Kaznacheev
and co-workers \cite{Bogdanov,Kaznacheev}, we assume the director
field ${\mathbf n}$ to be twisted bispherical, that is, the
director field lines connect virtual point defects \cite{Rudnick},
situated on the main-body axis of the drop and a distance
$2\widetilde{R}$ apart. (See also our previous work
\cite{Prinsen2}.) Each director field line lies on a surface of
revolution of a circle section about the main axis as shown in
Figure 1. Without loss of generality, we assume the latter to
coincide with the z-axis, and the origin to be separated a
distance $\widetilde{R}$ from both virtual defects. As for the
shape of the droplet, contrary to Williams' assumption, we assume
it not to be necessarily spherical but potentially elongated. Its
surface is the surface of revolution of a circle section around
the z-axis. This shape very accurately describes the shape of
actual tactoids \cite{Zocher,Prinsen2,Kaznacheev}. The distance
between the two poles of the droplet is set equal to $2R$, where
$R\leq\widetilde{R}$, and the droplet is positioned symmetrically
in the director field, that is, both poles are a distance $R$ away
from the origin (see Figure 2). We perform all the calculations in
bispherical coordinates $(\xi,\eta,\phi)$, because of the symmetry
of the problem in hand and because of the fact that a surface of
constant $\eta$ is the surface of revolution of a circle section,
see Figure 3. For further details of our parametrization see the
Appendix A.

We expect the twist deformation, if there is one, to be small near
the line connecting the two poles of the droplet and to grow
larger towards the surface of the droplet. The reason for this is
that introducing twist near the center of the droplet will greatly
increase the amount of bend deformation, whereas introducing it
near the surface has a much smaller effect. Hence we expect the
angle $\alpha$ between a director field line and a meridian on the
surface of revolution on which the line resides to depend strongly
on the $\eta$-coordinate but much more weakly on the
$\xi$-coordinate. As a first approximation, we neglect the
dependence on $\xi$ and assume that $\alpha=\alpha(\eta)$. To keep
things simple, we furthermore presume $\alpha$ to only depend on a
single parameter, i.e., $\alpha(\eta)=\alpha_{0}f(\eta)$, where
$\alpha_{0}$ sets the maximum twist angle at the surface of a drop
and $f(\eta)$ is some function of $\eta$. To fix the latter, we
demand that $\alpha(\eta)\propto \eta^{\delta}$ if $\eta\downarrow
0$, with $\delta>0$. This requirement is needed to keep the
contributions from the twist and the bend to the elastic free
energy deformation finite, and confirms the earlier claim that
near the center of the droplet the twist should be small.
Following Williams \cite{Williams1}, we use the Ansatz
$\alpha(\eta)=\alpha_{0}\sin\eta$ that fulfills all requirements.

Now that we have established our parametrization of the problem,we
can find the optimal droplet shape and director-field
configuration by minimizing the free energy at a constant volume
with respect to the free parameters $\epsilon$, $\widetilde{R}$
and $\alpha_{0}$. Before doing this, we first make the free energy
dimensionless by dividing it by $\tau V^{2/3}$. The dimensionless
free energy $\widetilde{F}$ then reads
\begin{equation}
\widetilde{F}=\frac{F}{\tau V^{2/3}}=\widetilde{F}_{S}+\omega
v^{-1/3}(\widetilde{F}_{11}+\widetilde{F}_{22}+\widetilde{F}_{33}),
\label{free}
\end{equation}
where $v=V(\tau\omega/K_{11})^{3}$ is a dimensionless volume, and
$\widetilde{F}_{S}=F_{S}/\tau V^{2/3}$ and
$\widetilde{F}_{ii}=F_{ii}/K_{11}
V^{1/3}=\gamma_{ii}I_{ii}/V^{1/3}$ are dimensionless free energies
with $I_{ii}\equiv F_{ii}/K_{ii}$; $F_{ii}$ is the contribution to
$F_{E}$ associated with the constant $K_{ii}$. See the Appendix A.
The ratios $\gamma_{11}\equiv K_{11}/K_{11}=1$, $\gamma_{22}\equiv
K_{22}/K_{11}$ and $\gamma_{33}\equiv K_{33}/K_{11}$ measure the
deviation from the equal constant approximation. Because of the
scaling $F_{ii}\propto V^{1/3}$ and $F_{S}\propto V^{2/3}$,
$\widetilde{F}_{ii}$ and $\widetilde{F}_{S}$ do not depend on the
volume of the droplet, only on its shape and director field. The
free energy equation (\ref{free}) is a function of the free
parameters $\omega$, $v$, $\gamma_{22}$ and $\gamma_{33}$.

\section{Bipolar drops: analytical results}

As already mentioned, Williams' criterion \cite{Williams1} for the
onset of the twist transition equation (\ref{wil}) is strictly
valid for spherical bipolar droplets only, i.e., in the limit of
infinitely large droplets \cite{Prinsen1,Prinsen2}. We extend this
result to droplets of finite size. Before turning to the more
general case of quasi-bipolar director fields we first allow for
non-spherical droplet shapes given a truly bipolar director field,
which corresponds to the situation where
$\omega\rightarrow\infty$. For that case we can analytically
calculate a generalization of inequality (\ref{wil}). Note that
according to theoretical estimates $\omega\approx 1$ for lyotropic
systems, although experiments point at somewhat larger values
closer to ten \cite{Prinsen1,Prinsen2,Kaznacheev}.

Let us assume that the transition from an untwisted to a twisted
director field is continuous, that is, the twist angle
$\alpha_{0}$ changes continuously from zero to a non-zero value,
an assumption that will be supported by numerical results
discussed below. Near the transition, we can then make a Taylor
expansion of the free energy for small $\alpha_{0}$,
\begin{equation}
\widetilde{F}\simeq\left.\widetilde{F}\right|_{\alpha=0}+
\left.\frac{\partial\widetilde{F}}{\partial\alpha}\right|_{\alpha=0}
\alpha_{0}+ \frac{1}{2}\left.\frac{\partial^{2}\widetilde{F}}
{\partial\alpha^{2}}\right|_{\alpha=0}\alpha_{0}^{2}.
\end{equation}
For symmetry reasons, the first non-constant term in $\alpha_{0}$
should be of second order in $\alpha_{0}$ so
$\partial\widetilde{F}/\partial\alpha=0$ for $\alpha=0$. At the
point where the twist transition occurs, the coefficient of the
$O(\alpha_{0}^{2})$-term vanishes. This coefficient can be
calculated by putting $R=\widetilde{R}$ (the director field being
bipolar) and expanding $(\nabla\cdot{\mathbf n})^{2}$, $({\mathbf
n}\cdot\nabla\times{\mathbf n})^{2}$ and $({\mathbf n}\times
(\nabla\times{\mathbf n}))^{2}$ in powers of $\alpha_{0}$, and
then performing the integrations of equation (\ref{frint}).
Because the droplet is bipolar, we have $({\mathbf n}\cdot{\mathbf
q})=0$ so the surface term $F_{S}$ does not depend on
$\alpha_{0}$.

We thus find for the coefficient of the $O(\alpha_{0}^{2})$-term
\begin{equation}
\left.\frac{1}{2}\frac{\partial^{2}\widetilde{
F}}{\partial\alpha^{2}} \right|_{\alpha=0}= \frac{4\pi\omega
R}{v^{1/3}V(\epsilon)^{1/3} (1+\epsilon^{2})^{2}}
\left(-4f_{1}(\epsilon)
+4f_{2}(\epsilon)\gamma_{22}+f_{3}(\epsilon) \gamma_{33} \right),
\label{twee}\end{equation} where the volume $V(\epsilon)$ is given
by equation (\ref{vol}), and
\begin{equation}
f_{1}(\epsilon)=(1+\epsilon^{2})^{2}\arctan^{2}\epsilon-\epsilon^{2},
\end{equation}
\begin{equation}
f_{2}(\epsilon)=f_{1}(\epsilon)+2\epsilon(1-\epsilon^{2})
\arctan\epsilon,
\end{equation}
\begin{equation}
f_{3}(\epsilon)=f_{1}(\epsilon)-2f_{2}(\epsilon)+4\epsilon^{2}.
\end{equation}
Bipolar droplets attain a twisted director-field structure, if
equation (\ref{twee}) becomes negative, i.e., if
\begin{equation}
f_{1}(\epsilon)\geq\gamma_{22}f_{2}(\epsilon)+
\frac{1}{4}\gamma_{33}f_{3}(\epsilon), \label{w2}
\end{equation}
because then the free energy is unstable to a perturbation away
from zero twist angle. Equation (\ref{w2}) is a generalization of
the Williams inequality equation (\ref{wil}) for elongated nematic
droplets of aspect ratio $\epsilon$. Setting $\epsilon=1$ in
equation (\ref{w2}) reproduces equation (\ref{wil}), as it should.
Figure 4 shows a contour plot of the reciprocal aspect ratio
$\epsilon$ as a function of $\gamma_{22}$ and $\gamma_{33}$ when
the equality in equation (\ref{w2}) holds. For a given value of
$\epsilon$, droplets of nematic liquid crystal with a combination
of elastic constants lying below the contour line corresponding to
that value of $\epsilon$ have a twisted structure, droplets lying
above are untwisted. From figure 4, we conclude that when the
aspect ratio increases ($\epsilon$ decreases), there are fewer
possible combinations of elastic constants that give a twisted
droplet structure, so the larger the aspect ratio, the harder it
is to get a twisted structure. Typical values of $\epsilon$
obtainable from experiments on colloidal dispersion are $0.1$ --
$0.5$ \cite{Sonin,Prinsen1,Kaznacheev}.

As it happens, the aspect ratio of a nematic drop is not a free
parameter---it sets itself so as to minimize the free energy. For
a bipolar droplet with $\widetilde{R}=R$ and $\alpha(\eta)\equiv
0$ we find that the free energy has the form (see also \cite{Prinsen1}
\begin{equation}\fl
\widetilde{F}=8\pi (1+\epsilon^{2}) c_{2}(\epsilon)
c_{3}(\epsilon)^{2}+8\pi\omega v^{-1/3}c_{3}(\epsilon)
\left(\left(1+\frac{3}{4}\gamma_{33}\right)c_{2}(\epsilon)-
\gamma_{33}\epsilon\arctan^{2}\epsilon\right),\label{fr}
\end{equation}
where
\begin{equation}
c_{2}(\epsilon)=\epsilon-(1-\epsilon^{2})\arctan\epsilon
\end{equation}
and
\begin{equation}
c_{3}(\epsilon)=\left(\frac{3}{4\pi}\right)^{1/3}
\left(3(1+\epsilon^{2})^{2}c_{2}(\epsilon)-4\epsilon^{3}\right)^{-1/3}.
\end{equation}
This free energy is valid all the way to the critical conditions
set by equation (\ref{w2}). To find the optimal $\epsilon$ for a
given droplet volume $v$ and elastic constants $\gamma_{22}$ and
$\gamma_{33}$, we simply set $\partial\widetilde{F}
/\partial\epsilon=0$. This produces an implicit expression for
$\epsilon$ that we do not reproduce here. We have only been able
to find an explicit solution if $v^{1/5}\omega^{-3/5}\lesssim 1$,
in which case $\epsilon\sim (1/4)(15/8\pi)^{1/5}
v^{1/5}\omega^{-3/5}\ll 1$ and the drops are very elongated. For
small $\epsilon$, $f_{1}(\epsilon)\sim 4\epsilon^{4}/3$,
$f_{2}(\epsilon)\sim 2\epsilon^{2}$ and $f_{3}(\epsilon)\sim
4\epsilon^{4}$, so the parity broken structure only occurs if
$\gamma_{22}$ is very small, i.e., at most $\epsilon^{2}$ or
smaller than that, unless $\gamma_{33} \geq 4/3$ in which case the
twisted state is absolutely unstable for all $\gamma_{22}$. This
confirms again that twisted director fields are very unlikely in
elongated tactoids.

We note that since $f_{2}(\epsilon)/f_{1}(\epsilon)$ and
$f_{3}(\epsilon)/f_{1}(\epsilon)$ are both monotonically
decreasing functions of $\epsilon$, and since $\epsilon$ is a
monotonically increasing function of the droplet volume
\cite{Prinsen1}, that if a twist transition occurs in a bipolar droplet
at $v=v_{0}$ the droplet must be twisted for all $v>v_{0}$
but untwisted for $v<v_{0}$. Our numerical studies, presented in
the next section confirm this.

Returning to the issue whether or not the saddle-splay elastic
deformation has any appreciable impact on the twist transition, we
find that if we add its contribution
\begin{equation}
-K_{24}\int _{V}d^{3}{\mathbf r}\hspace{2pt} \nabla \cdot \left[
{\mathbf n}\left( \nabla \cdot {\mathbf n}\right)+{\mathbf n}
\times \left( \nabla \times {\mathbf n}\right)\right]
\end{equation}
to the total free energy, the generalized Williams inequality, equation
(\ref{w2}), becomes
\begin{equation}
f_{1}(\epsilon)+\gamma_{24}f_{4}(\epsilon)\geq\gamma_{22}
f_{2}(\epsilon)+ \frac{1}{4}\gamma_{33}f_{3}(\epsilon). \label{w3}
\end{equation}
Here $\gamma_{24}\equiv K_{24}/K_{11}$ and $f_{4}(\epsilon) =
\epsilon (1-\epsilon^{2})\arctan\epsilon$. We immediately see that
for nearly spherical drops we retrieve the original Williams
inequality, (\ref{wil}), because $f_{4}(\epsilon)\rightarrow 0$ as
$\epsilon\uparrow 1$. This implies that the splay-bend
contribution is small compared to the other elastic contributions
for nearly spherical droplets and that we can indeed neglect it.
However, in the opposite limit where the droplet is strongly
elongated this in no longer the case. Inserting the limit
$\epsilon\downarrow 0$ in equation (\ref{w3}), our generalized
Williams inequality becomes $\gamma_{24}\geq 2\gamma_{22}$,
showing that the splay-twist elastic term may indeed stabilize the
twisted state if the splay-bend elastic constant is sufficiently
large. Unfortunately, as far as we are aware, not even an
order-of-magnitude estimate is known of $\gamma_{24}$, at least
not for lyotropic nematics. We can, in spite of this, nonetheless
make headway by presuming that the Nehring-Saupe equality
\cite{Nehring} $2\gamma_{24}= 1- \gamma_{22}$ holds. (See,
however, \cite{Yokoyama}.) Presuming it holds, our generalized
Williams inequality simplifies to $\gamma_{22}\leq 1/5$,
indicating that very slender tactoids formed in dispersions of
semi-flexible polymers could potentially undergo a parity-breaking
transition induced by the splay-bend surface elasticity. We think
that this phenomenon is unlikely to be observed in practice,
because typically $\epsilon > 0.1$ \cite{Sonin,Prinsen1,Bogdanov},
i.e., tactoids are in practice just not sufficiently slender. In
addition, to become sufficiently slender they would have to be
very small, and small tactoids tend to revert to a uniform
director field as the calculations presented in the next section
show. For these reasons, we think it justified to focus on the
bulk elasticity of the droplets.

\section{General case: numerical results}

Now we consider the more general case, where the director field in
the droplet is intermediate between uniform and truly bipolar. As
we \cite{Prinsen2}, and others \cite{Kaznacheev} recently argued,
nematic tactoids are usually only approximately bipolar, with a
director field that becomes increasingly uniform the smaller their
size. For this more general case it is not possible to
analytically derive a further generalization of equation
(\ref{wil}). We therefore proceed to investigate the droplet shape
and structure using numerical methods. The optimal values for the
aspect ratio $\epsilon$, degree of curvature of the director field
$\widetilde{R}$ and maximum twist angle $\alpha_{0}$ are
determined by numerically minimizing equation (\ref{free}) with
respect to these quantities at fixed values of $\omega$, $v$,
$\gamma_{22}$ and $\gamma_{33}$, using the quasi-Newton algorithm
E04JYF from the NAG$^{\circledR}$ Fortran Library, release Mark
18.

As already mentioned, the free energy cost of introducing a twist
in the director field near the surface of the droplet is less than
that in the center. This implies that introducing twist near the
surface of a spherical droplet is less costly in terms of free
energy than near the surface of an elongated droplet, assuming
$\widetilde{R}/R$ is the same for both droplets. We also expect
that introducing twist in a bipolar droplet, where there is a lot
of splay deformation near the poles, has a larger effect than
introducing twist in a droplet of the same shape, but with a more
homogeneous director field.

This then leads us to the conjecture that if  a transition occurs
for a certain value $v=v_{0}$ for a given set of fixed values of
$\omega$, $\gamma_{22}$ and $\gamma_{33}$, the droplet must be
twisted for all $v>v_{0}$ but untwisted for $v<v_{0}$. This is
true for bipolar droplets as we showed above, but is plausible for
the more general case too on account of how the aspect ratio
$\epsilon^{-1}$ depends on $v$. (See, again, reference
\cite{Prinsen2}.) One of the consequences of this conjecture is
that if inequality (\ref{wil}) predicts a droplet to be untwisted,
the director field inside the droplet will be untwisted at every
volume. The reason is that equation (\ref{wil}) is formally valid
only for $v\rightarrow\infty$.

The first case we investigate is the so called equal-constant
approximation $\gamma_{22}=\gamma_{33}=1$, for which
$K_{11}=K_{22}=K_{33}$. This combination of elastic constants does
not satisfy inequality (\ref{wil}), so, on account of the
conjecture we put forward in the previous paragraph, we do not
expect a twisted director field in the droplet at any volume. This
is indeed what we find from the numerical minimization: for
$\omega=0.01$, $1$ and $100$, $\alpha_{0}=0$ for all $v$ tested in
the range from $10^{-18}$ to $10^{18}$. Next, we study a
combination of elastic constants valid for rod-like colloids.
Putting $\gamma_{22}=1/3$ in equation (\ref{wil}), we see that for
a spherical bipolar droplet to be twisted, we must have
$\gamma_{33}<1.55$, which is much lower than the typical value for
rod-like colloids of $\gamma_{33}\gtrapprox 5$. (See section 2.)
So, again from the conjecture from the previous paragraph, we do
not expect a twisted director field at any droplet volume. This is
confirmed for $\gamma_{33}=5$ and $\omega=0.01$, $1$ and $100$ by
a numerical minimization for dimensionless volumes from $10^{-18}$
to $10^{18}$.

Let us now keep the value of $\gamma_{22}$ fixed at $1/3$, but
take for $\gamma_{33}$ a value of $1$. Inequality (\ref{wil}) now
predicts a twisted director field to be stable in the limit of an
infinite volume implying that it might also be stable for some
finite volume. That this is indeed the case, is shown in Figure 5,
where we show the twist angle $\alpha_{0}$ as a function of the
dimensionless volume $v$ for $\omega=0.01$, $1$ and $100$. We see
that the transition is sharp and continuous, which was our
assumption in deriving the results of section 3. This means that
the droplet is absolutely untwisted until the volume reaches the
critical size $v=v_{0}$, from whereon the twist gradually
increases with increasing $v$. For $\omega=100$, the twist
transition is not shown, but according to our calculations it
occurs at $^{10}\log v_{0}=10.60$. This can in fact be checked
numerically via a different route. The aspect ratio at the
transition can be calculated using the generalized Williams
condition equation (\ref{w2}). Because the director field at the
transition is nearly bipolar (see figure 6 and the next
paragraph), the aspect ratio should also minimize equation
(\ref{fr}), which then determines $v_{0}$. We find that $^{10}\log
v_{0}=10.61$ in good agreement with our previous estimate.

In figure 6 we have plotted a diagram of states for
$\gamma_{22}=1/3$ and $\gamma_{33}=1$. The twist transition is the
only transition in the diagram that is sharp, i.e., on one side of
the line demarcating the transition, the droplets are untwisted
whereas they are twisted on the other. The line demarcating the
transition from a homogeneous to a bispherical director field is
drawn at the point where $\widetilde{R}=2R$, and one showing the
transition in the aspect ratio from spheroidal to elongated is set
at $\epsilon=1/2$. Both are smooth crossovers, discussed in more
detail in reference \cite{Prinsen2}. This means that a droplet,
e.g., in that part of the phase diagram indicated by "spheroidal
twisted bipolar" is not exactly spherical nor exactly bipolar, but
nearly so albeit that the director field is indeed twisted.

Figure 6 shows that, at the twist transition, $v$ is independent
of $\omega$ when $\omega$ is small, whereas for large values of
$\omega$ this is not so with $v\sim \omega^{3}$. The first result
can easily be understood by noting that $\omega$ describes the
degree of anisotropy of the anisotropic surface tension. If
$\omega$ is small, this anisotropy is small and it will not
influence the droplet shape nor the director-field configuration.
The second observation follows from the fact that when the droplet
is bipolar, the aspect ratio must be constant at the twist
transition. This can be seen from equation (\ref{w2}), which is
independent of $\omega$, implicating that the aspect ratio is
constant as a function of $\omega$ if $\gamma_{22}$ and
$\gamma_{33}$ are constant. We have shown in previous work
\cite{Prinsen2} that if $\omega\gg 1$, the aspect ratio $\epsilon$
of a quasi-bipolar droplet scales as $v^{-1/5}\omega^{3/5}$, from
which one can conclude that $v\sim \omega^{3}$ if $\epsilon$ is
constant and $\omega\gg 1$. (See also equation (\ref{fr}).)

Next, we take $\gamma_{22}=0.05$ and $\gamma_{33}=1$, which are
representative values of the elastic constants for semi-flexible
particles (see section 2). In figures 7 and 8, we show the twist
angle $\alpha_{0}$ as a function of the dimensionless volume $v$
for anchoring strengths $\omega=0.01$, $1$ and $100$, and the
corresponding diagram of states. The behavior of the droplets is
similar as for the previous set of parameter values, except that
the twist transition now occurs at a much lower value of $v$ and
that the diagram of states is somewhat more complex. This is to be
expected, considering that the twist elastic constant is much
smaller. The twist transition for a strong anchoring with
$\omega=100$ is not shown, but it occurs at $^{10}\log
v_{0}=7.71$. Using the same method as described earlier in this
section we can check this result with the theory of section 4. We
find $^{10}\log v_{0}=7.79$. The discrepancy between the two
results is probably caused by the fact that the droplet is not
quite bipolar yet (see figure 8).

As already mentioned, we expect $\omega\approx 1$ to be a
reasonable estimate for the dimensionless anchoring strength. (See
again, however, \cite{Prinsen1,Prinsen2,Kaznacheev}.) Because the
elastic constants and their ratios differ from system to system,
we show, in Figure 9, the dimensionless volume $v_{0}$ at which
the twist transition occurs for various values of $\gamma_{22}$
and $\gamma_{33}$ at a fixed anchoring strength $\omega=1$. As
shown in the figure, the larger $\gamma_{33}$, the larger the
volume of a tactoid needs to be to be able to support a twisted
director field. The reason is that a twist deformation goes hand
in hand with a bend deformation that is penalized more the larger
the bend elastic constant $K_{33}$. Also, the larger
$\gamma_{22}$, the larger the elastic free energy penalty
associated with a twist deformation, the smaller $\gamma_{33}$
must be for the twisted director field to be favorable. The shift
to higher droplet volumes is, as discussed, linked with the
greater ease of a twist deformation the larger the drop is because
large drops are more spherical than small ones. It is useful to
point out that provided the values of the elastic constants and
the surface tension are known, figure 9 can be used to estimate
the minimum droplet volume at which a transition to a twisted
structure should be expected to occur.

We again turn to the question why twisted nematic droplets have so far
evaded observation in lyotropic liquid crystals whereas they have
been observed in thermotropic ones \cite{Drzaic1,Volovik}. We have
seen that we should not expect twisted nematic droplets in
nematics of hard rods because $\gamma_{22}$ and $\gamma_{33}$ are
too large. For worm-like, semi-flexible particles such as fd virus
though, $\gamma_{22}$ and $\gamma_{33}$ possibly have the right
values but the volume of the droplets has to be sufficiently
large for parity breaking to occur.

In fact, we can use figure 9 to estimate how large. First,
we know that $v\propto V\tau^{3}$. Since the elastic constants of
a thermotropic and a lyotropic crystal are of the same order of
magnitude, we expect the twist transition to occur at about the
same value of $v$. However, $\tau$ is typically one or two orders
of magnitude larger in a thermotropic liquid crystal than in a
lyotropic one
\cite{Williams2,Chen1,Chen2,Schoot,Lang,Chen6,Chen3,Chen4,Chen5}.
This, then, means that the volume of a droplet of lyotropic liquid
crystal at the twist transition must be three to six orders of
magnitude larger than that of a thermotropic liquid crystal.

To make our estimate more concrete, we set
$K_{33}/K_{11}=1$, $K_{22}/K_{11}=0.05$ and $\omega=1$, and read off
from figure 9 that the twist transition occurs if
$v_{0}=V(\tau\omega/K_{11})^{3}=10^{1.35}$. Inserting the typical values
$K_{11}=10^{-11}$ N and $\tau=10^{-6}$ N/m, this corresponds to a droplet
volume of $V\approx 10^{4}$ $\mu$m$^{3}$, with a linear
size of about $30$ $\mu$m. However, if $\omega$ is closer to $10$,
as might well be the case \cite{Prinsen1,Prinsen2,Kaznacheev},
this value increases to $>300$ $\mu$m. A similar enhancement
occurs if $\tau$ is larger by a factor of ten.

\section{Conclusion}

By adopting a twisted director field structure, a nematic tactoid
reduces its splay energy albeit at the expense of increasing its
bend and twist energy. Hence, if the bend and twist elastic
constants are small enough in comparison to the splay elastic
constant, this twisted (parity-broken) structure is energetically
more favorable. For a spherical bipolar droplet, Williams
\cite{Williams1} derived a criterion, equation (\ref{wil}), the
elastic constants have to obey in order to be able to support a
twisted director field. In this work we generalized this criterion
to bipolar droplets of arbitrary aspect ratio, equation
(\ref{w2}), showing that the more elongated the drop the less
likely they are to have a twisted director field.

We find the transition to a twisted structure to be sharp, in
contrast to the transformation from a uniform director field to a
bipolar one. When the transition occurs at a certain dimensionless
droplet volume $v=v_{0}$, the droplet is twisted for all $v>v_{0}$
but untwisted for $v<v_{0}$. One of the consequences is that
droplets that do not have a twisted structure at infinite volume
(the limit in which Williams' result is valid), cannot have a
twisted structure for any finite volume either. Not entirely
surprisingly, $v_{0}$ increases with increasing $\gamma_{33}\equiv
K_{33}/K_{11}$ and decreases with increasing $\gamma_{22}\equiv
K_{22}/K_{11}$.

Nematics of rod-like colloidal particles appear to have elastic
constants that do not obey the Williams inequality, so we do not
expect to find droplets with a twisted structure in this sort of
system. This is not so for dispersions of semi-flexible colloidal
particles (or polymers), however, but the tactoids have to be
large enough for the twisted director-field configurations to
become stable. This is in part due to them becoming less elongated
with increasing size.

\ack

We are grateful to P. Teixeira for drawing our attention to
reference \cite{Yokoyama}.

\appendix
\section*{Appendix A}
\setcounter{section}{1}

The bispherical coordinates are related to the Cartesian
coordinates ${\mathbf x}=(x,y,z)$, through the relations
\begin{equation}
\left\{
\begin{array}{rcl}
x & = & RZ^{-1}\sin\eta\sin\xi\cos\phi, \vspace{8pt}\\
y & = & -RZ^{-1}\sin\eta\sin\xi\sin\phi, \vspace{8pt}\\
z & = & RZ^{-1}\cos\xi,
\end{array}
\right.
\end{equation}
where $Z=1+\sin\xi\cos\eta$ and $0\leq\xi\leq\pi$,
$0\leq\eta\leq\pi$ and $0\leq\phi<2\pi$. The Cartesian unit
vectors ${\mathbf e}_{x}$, ${\mathbf e}_{y}$ and ${\mathbf e}_{z}$
are related to the bispherical unit vectors ${\mathbf e}_{\xi}$,
${\mathbf e}_{\eta}$ and ${\mathbf e}_{\phi}$ by
\begin{equation}\fl
\left\{
\begin{array}{rcl}
{\mathbf e}_{x} & = &
Z^{-1}\sin\eta\cos\xi\cos\phi\hspace{3pt}{\mathbf e}_{\xi}+
Z^{-1}(\sin\xi+\cos\eta)\cos\phi\hspace{3pt}{\mathbf e}_{\eta}-
\sin\phi\hspace{3pt}{\mathbf e}_{\phi}, \vspace{8pt}\\
{\mathbf e}_{y} & = &
-Z^{-1}\sin\eta\cos\xi\sin\phi\hspace{3pt}{\mathbf e}_{\xi}-
Z^{-1}(\sin\xi+\cos\eta)\sin\phi\hspace{3pt}{\mathbf e}_{\eta}-
\cos\phi\hspace{3pt}{\mathbf e}_{\phi}, \vspace{8pt}\\
{\mathbf e}_{z} & = &
-Z^{-1}(\sin\xi+\cos\eta)\hspace{3pt}{\mathbf e}_{\xi}+
Z^{-1}\sin\eta\cos\xi\hspace{3pt}{\mathbf e}_{\eta}.
\end{array}
\right.
\end{equation}
Finally, $h_{\xi}=|\partial{\mathbf x}/\partial\xi|=RZ^{-1}$,
$h_{\eta}=|\partial{\mathbf x}/\partial\eta|=RZ^{-1}\sin\xi$ and
$h_{\phi}=|\partial{\mathbf
x}/\partial\phi|=RZ^{-1}\sin\xi\sin\eta$ are the metric components
of the bispherical coordinate system.

The director field ${\mathbf n}$ is given by
\begin{equation}\fl
{\mathbf
n}(\xi,\eta,\phi)=-\frac{d_{2}}{\sqrt{d_{1}^{2}+d_{2}^{2}}}
\cos\alpha(\eta){\mathbf
e}_{\xi}+\frac{d_{1}}{\sqrt{d_{1}^{2}+d_{2}^{2}}}
\cos\alpha(\eta){\mathbf e}_{\eta}+\sin\alpha(\eta){\mathbf
e}_{\phi},
\end{equation}
where
\begin{equation}
d_{1} = (\widetilde{R}^{2}-R^{2})Z\sin\eta\cos\xi,
\end{equation}
and
\begin{equation}
d_{2} =
(\widetilde{R}^{2}-R^{2})Z\cos\eta+(\widetilde{R}^{2}+R^{2})Z\sin\xi.
\end{equation}
This choice of director field leads to the following expression
for the various components of the elastic free energy, equation
(\ref{blk}),
\begin{equation}
\nabla\cdot{\mathbf n}=\frac{Z}{\sqrt{d_{1}^{2}+d_{2}^{2}}}
(4\cos\xi\cos\alpha+d_{1}\alpha_{\eta}\sin\alpha),\label{div}
\end{equation}
\begin{equation}
{\mathbf n}\cdot\nabla\times{\mathbf n}=
\frac{-Z}{\sin\xi\sqrt{d_{1}^{2}+d_{2}^{2}}}
\left(\frac{\sin2\alpha}{\sin\eta}Z\left(\frac{\widetilde{R}^{2}+1}{2}
Z-1\right)+d_{2}\alpha_{\eta}\right),
\end{equation}
\begin{equation}\fl
({\mathbf n}\times(\nabla\times{\mathbf n}))_{\xi}=
\cot\xi\sin^{2}\alpha+\frac{Zd_{1}\cos\alpha}
{\sin\xi(d_{1}^{2}+d_{2}^{2})}(2\sin^{2}\xi\sin\eta\cos\alpha-
d_{2}\alpha_{\eta}\sin\alpha),
\end{equation}
\begin{equation}\fl
({\mathbf n}\times(\nabla\times{\mathbf n}))_{\eta}=
\frac{\sin^{2}\alpha(\sin\xi+\cos\eta)}{\sin\xi\sin\eta}+
\frac{d_{1}^{2}Z\alpha_{\eta}\sin2\alpha}{2\sin\xi(d_{1}^{2}+d_{2}^{2})}+
\frac{2d_{2}Z\sin\xi\sin\eta\cos^{2}\alpha}{d_{1}^{2}+d_{2}^{2}},
\end{equation}
and
\begin{equation}\fl
({\mathbf n}\times(\nabla\times{\mathbf n}))_{\phi}=
\frac{-\cos\alpha}{\sqrt{d_{1}^{2}+d_{2}^{2}}}
\left(\frac{d_{1}\sin\alpha(\sin\xi+\cos\eta)}{\sin\xi\sin\eta}+
\frac{d_{1}Z\alpha_{\eta}\cos\alpha}{\sin\xi}-d_{2}\cot\xi\sin\alpha\right),\label{rot3}
\end{equation}
where $\alpha_{\eta}=d\alpha(\eta)/d\eta$ and $({\mathbf
n}\times(\nabla\times{\mathbf n}))_{i}$ denotes the component of
${\mathbf n}\times(\nabla\times{\mathbf n})$ in the $i$-direction,
where $i\in \{\xi,\eta,\phi\}$.

Within the bipolar coordinate system, the surface of a droplet
with aspect ratio $\epsilon^{-1}$ is given by
$\eta=\eta_{0}=2\arctan\epsilon$, $0\leq\xi\leq\pi$ and
$0\leq\phi<2\pi$. So, using equations (\ref{div}) -- (\ref{rot3}),
we can calculate the splay, twist and bend contributions
$F_{11}\equiv K_{11}I_{11}$, $F_{22}\equiv K_{22}I_{22}$ and
$F_{33}\equiv K_{33}I_{33}$ to the Frank elastic energy,
$F_{E}=F_{11}+F_{22}+F_{33}$ of equation (\ref{blk}), where
\begin{equation}\fl
\begin{array}{l}
I_{11} = \frac{1}{2}\int _{V}d^{3}{\mathbf r}\left(
\nabla\cdot{\mathbf n}\right)^{2}=\frac{1}{2}\int
_{0}^{\pi}d\eta\int _{0}^{\eta_{0}}d\xi\int _{0}^{2\pi}d\phi
h_{\xi}h_{\eta}h_{\phi}\left( \nabla\cdot{\mathbf
n}\right)^{2}, \vspace{8 pt}\\
I_{22} = \frac{1}{2}\int _{V}d^{3}{\mathbf r}\left( {\mathbf
n}\cdot \nabla\times{\mathbf n}\right)^{2}=\frac{1}{2}\int
_{0}^{\pi}d\eta\int _{0}^{\eta_{0}}d\xi\int _{0}^{2\pi}d\phi
h_{\xi} h_{\eta}h_{\phi}\left(
{\mathbf n}\cdot \nabla\times{\mathbf n}\right) ^{2}, \vspace{8 pt}\\
I_{33} = \frac{1}{2}\int _{V}d^{3}{\mathbf r} \left( {\mathbf
n}\times\left( \nabla\times{\mathbf
n}\right)\right)^{2}=\frac{1}{2}\int _{0}^{\pi}d\eta\int
_{0}^{\eta_{0}}d\xi\int _{0}^{2\pi}d\phi h_{\xi}
h_{\eta}h_{\phi}\left( {\mathbf n}\times
\left(\nabla\times{\mathbf n}\right)\right)^{2}.
\end{array}
\label{frint}
\end{equation}
The surface free energy becomes
\begin{equation}\fl
F_{S}=\tau\int_{S}d^{2}{\mathbf r}(1+\omega({\mathbf
n}\cdot{\mathbf q})^{2})
=\tau\int_{0}^{\pi}d\xi\int_{0}^{2\pi}d\phi (1+\omega({\mathbf
n}\cdot{\mathbf q})^{2})h_{\xi}h_{\phi}|_{\eta=\eta_{0}},
\end{equation}
with, for our choice of director field and droplet surface,
\begin{equation}
({\mathbf n}\cdot{\mathbf
q})^{2}=\frac{d_{1}^{2}}{d_{1}^{2}+d_{2}^{2}}\cos^{2}\alpha.
\end{equation}
For the volume $V$ and the surface area $S$ of the droplet we find
\begin{equation}\fl
V(\epsilon)=\int _{0}^{\pi}d\eta\int _{0}^{\eta_{0}}d\xi\int
_{0}^{2\pi}d\phi h_{\xi}h_{\eta}h_{\phi} = \frac{\pi }{6}
R^{3}\left[ 3\left( \frac{1+\epsilon^{2}}{\epsilon}\right)^{2}
\left(1- \frac{1-\epsilon^{2}}{\epsilon}\arctan
\epsilon\right)-4\right]\label{vol}
\end{equation}
and
\begin{equation}
S(\epsilon)=\int _{0}^{\pi}d\xi\int _{0}^{2\pi}d\phi
h_{\xi}h_{\phi}|_{\eta=\eta_{0}} = 2\pi
R^{2}\frac{1+\epsilon^{2}}{\epsilon}
\left[1-\frac{1-\epsilon^{2}}{\epsilon} \arctan \epsilon\right].
\end{equation}

\newpage

\section*{Figure Captions}

$ $

{\bf Figure 1.} Sketch of a twisted director-field line in a
spindle-like nematic droplet or tactoid. The twist is largest near
the surface of the drop. Close to the center line of the drop the
twist is presumably negligible.

{\bf Figure 2.} Droplet surface (thick lines) and defining
surfaces for the director field lines (thin lines). The system is
rotationally symmetric around the line connecting the two
(virtual) defects given by the crossing or focal points of the
director field lines. $R$ is half the length of the major axis of
the droplet. $\widetilde{R}$ is half the distance between the
virtual defects.

{\bf Figure 3.} The bispherical coordinate system. The unit vector
${\mathbf e}_{\phi}$ points out of the plane on the right hand
side of the line indicated by $\eta=0$. The picture should be
rotated along the line $\eta=0$ to get the full three dimensional
coordinate system. See the main text for an explanation of the
symbols.

{\bf Figure 4.} Contour plot of the inverse of the aspect ratio
$\epsilon$ of a tactoid exactly at the twist transition as a
function of the ratios $\gamma_{33}=K_{33}/K_{11}$ and
$\gamma_{22}=K_{22}/K_{11}$ for bipolar droplets where $K_{33}$,
$K_{22}$ and $K_{11}$ are the usual bend, twist and splay elastic
constants. The numbers next to the contour lines denote the values
of $\epsilon$. A bipolar droplet has a twisted structure if and
only if the point denoting the combination of the ratios of
elastic constants $\gamma_{22}$ and $\gamma_{33}$ of the nematic
lies below the contour line corresponding to the aspect ratio of
the droplet.

{\bf Figure 5.} The maximum (surface) twist angle $\alpha_{0}$ as
a function of the dimensionless volume $v$ in units of $\pi$
radians for the ratios of the elastic constants
$\gamma_{33}=K_{33}/K_{11}=1$ and $\gamma_{22}=K_{22}/K_{11}=1/3$
at three different values of the anchoring strength $\omega$. For
$\omega=100$, the twist transition occurs at $^{10}\log v=10.60$
(not shown).

{\bf Figure 6.} The diagram of states for
$\gamma_{33}=K_{33}/K_{11}=1$ and $\gamma_{22}=K_{22}/K_{11}=1/3$.
The diagram shows the shape of the droplet as a function of the
dimensionless volume $v$ and the anchoring strength $\omega$.
Indicated are the transition from a uniform director field to a
bipolar one (vertical line), the transition from a spheroidal to
an elongated shape (horizontal line) and the transition from a
bipolar to a twisted director field (line in the lower right
corner). Only the latter transition is sharp, the transition from
uniform to bipolar is (arbitrarily) taken at the boojum position
$\widetilde{R}/R=2$ and the transition from spheroidal to bipolar
at an aspect ratio of 2.

{\bf Figure 7.} The twist angle $\alpha_{0}$ as a function of the
dimensionless volume $v$ in units of $\pi$ radians for
$\gamma_{33}=K_{33}/K_{11}=1$ and $\gamma_{22}=K_{22}/K_{11}=0.05$
at three different values of the anchoring strength $\omega$.  For
$\omega=100$, the twist transition occurs at $^{10}\log v=7.71$
(not shown).

{\bf Figure 8.} As figure 6, but now with
$\gamma_{33}=K_{33}/K_{11}=1$ and
$\gamma_{22}=K_{22}/K_{11}=0.05$. The various director-field
patterns and droplet shapes are again indicated. Notice the
increased number of regimes.

{\bf Figure 9.} The value of $v$ at which the twist transition
occurs as a function of $\gamma_{33}$ for various values of
$\gamma_{22}\equiv K_{22}/K_{11}$ and with $\omega=1$. For given
$\gamma_{22}$, the droplets above the line have an untwisted
structure, whereas the droplets below it have a twisted structure.
By adopting a twisted structure, a droplet decreases its splay
energy, at the same time increasing its twist and bend energy. For
that reason, the twisted structure occurs only if
$\gamma_{33}\equiv K_{33}/K_{11}$ is small enough, even when
$\gamma_{22}=0$.

\vspace{5 cm}

\section*{Figures}

\begin{figure}[b]
\begin{center}
\leavevmode \epsfig{file=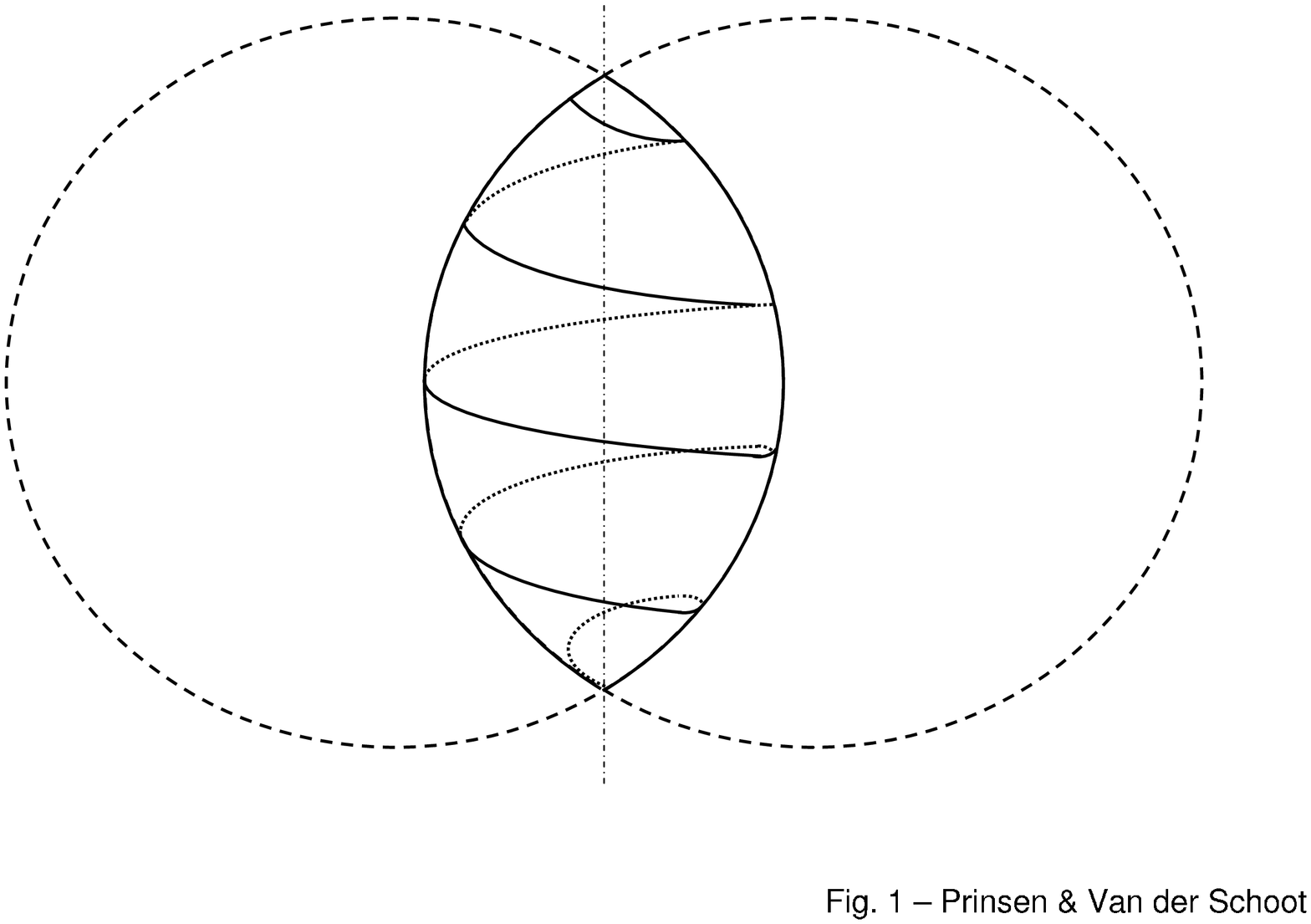, angle=-90, totalheight=11 cm}
\end{center}
\end{figure}

\newpage

\begin{figure}
\begin{center}
\leavevmode \epsfig{file=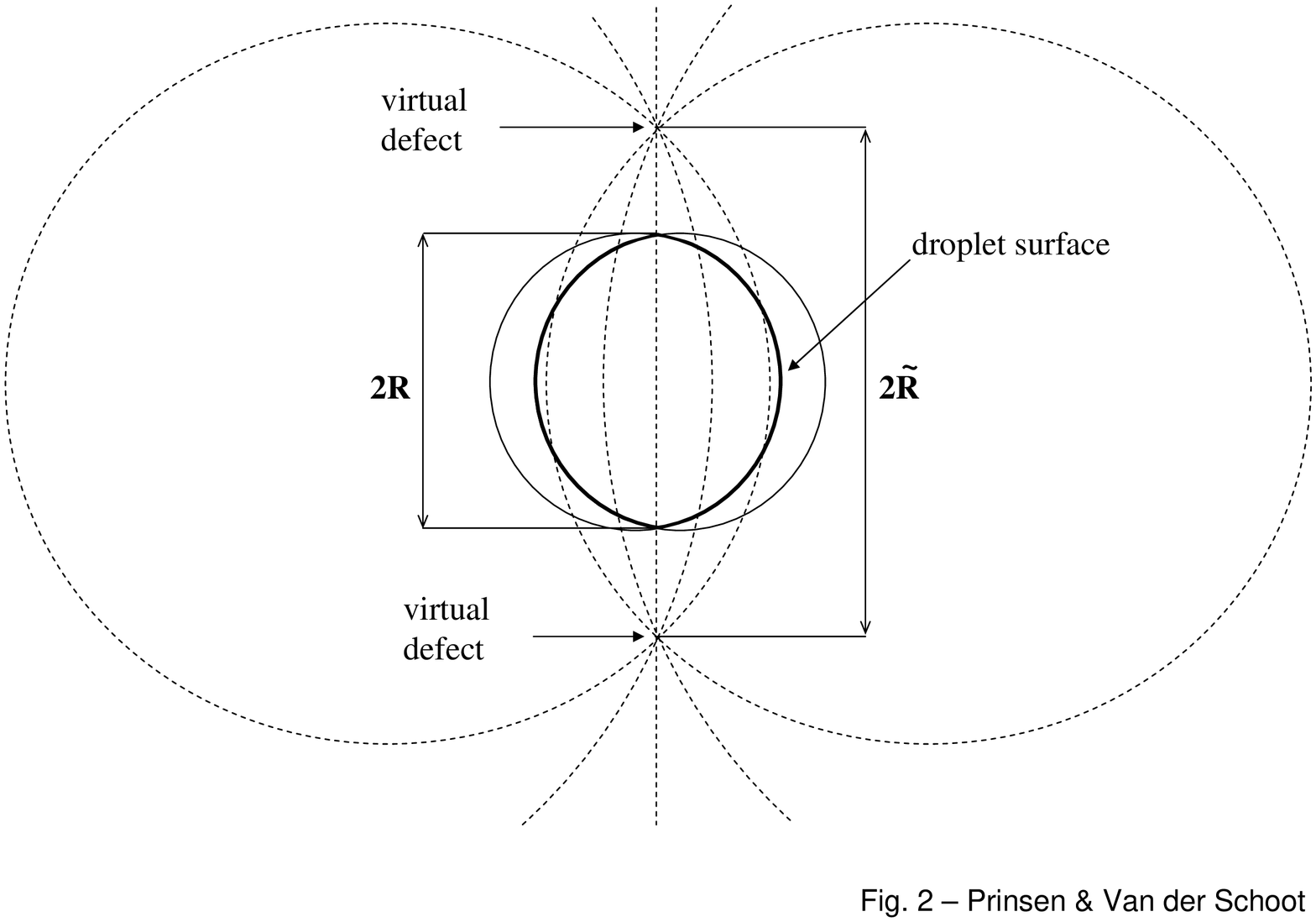, angle=-90, totalheight=11 cm}
\end{center}
\end{figure}

\begin{figure}
\begin{center}
\leavevmode \epsfig{file=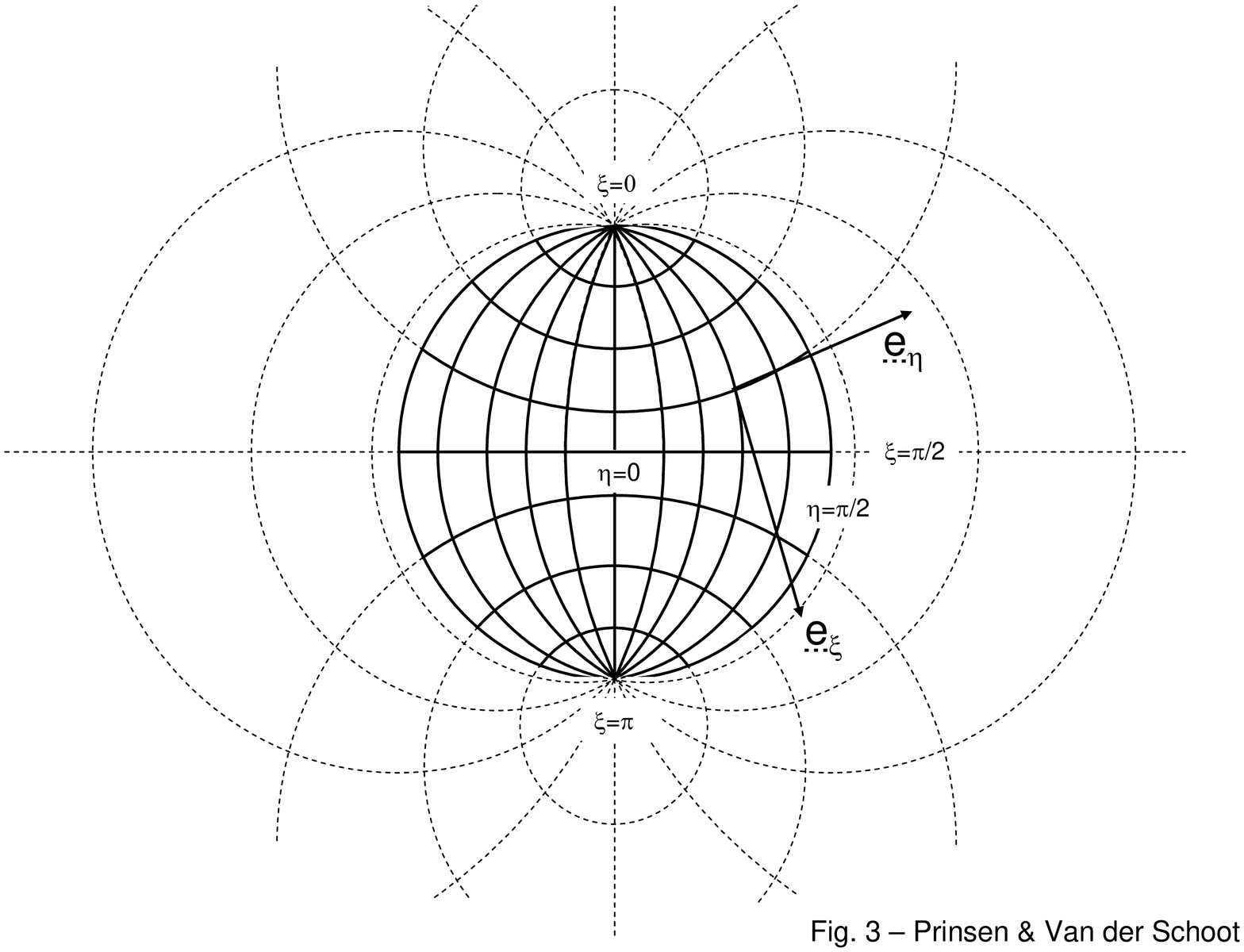, angle=-90, totalheight=11 cm}
\end{center}
\end{figure}

\begin{figure}
\begin{center}
\leavevmode \epsfig{file=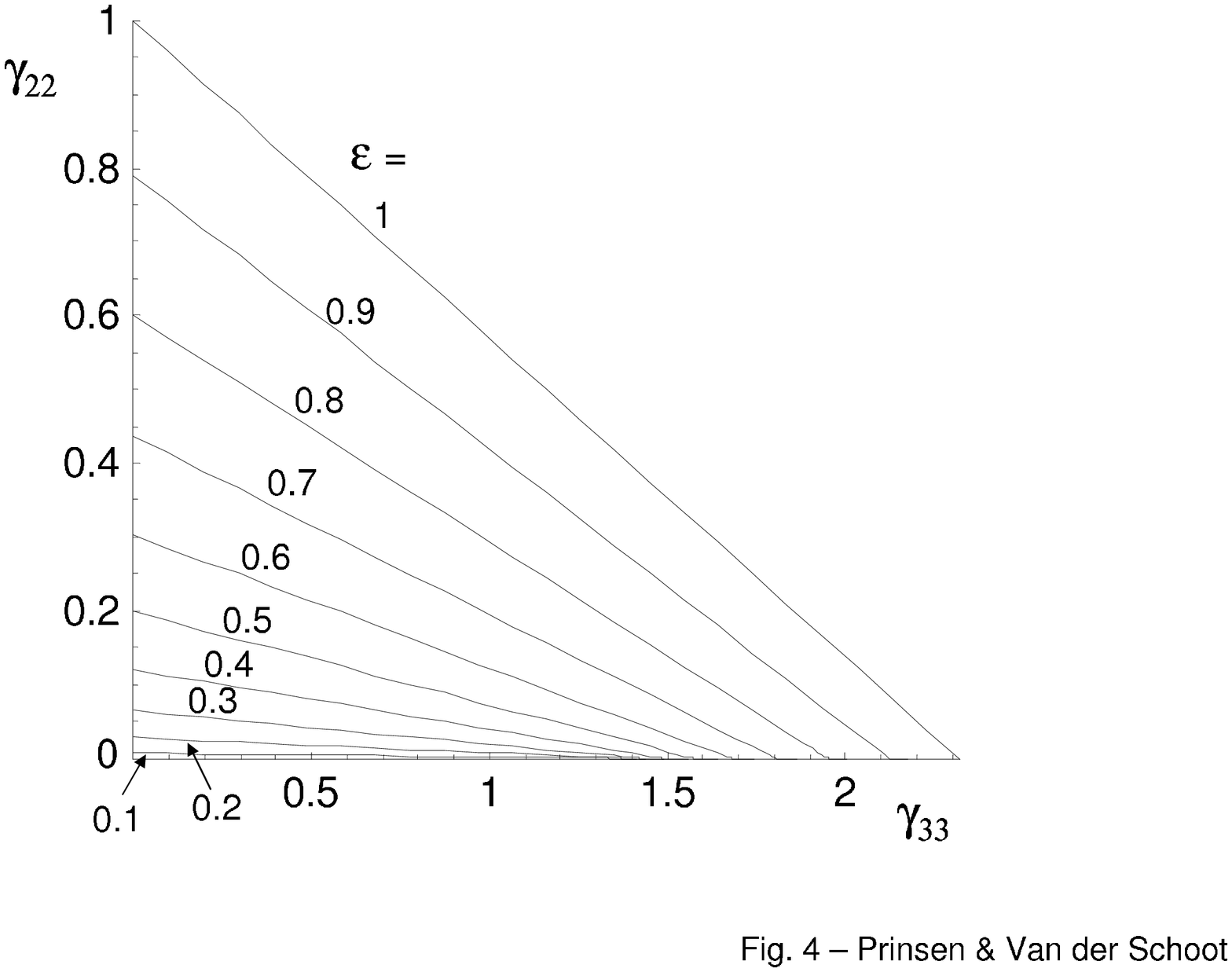, angle=-90, totalheight=11 cm}
\end{center}
\end{figure}

\begin{figure}
\begin{center}
\leavevmode \epsfig{file=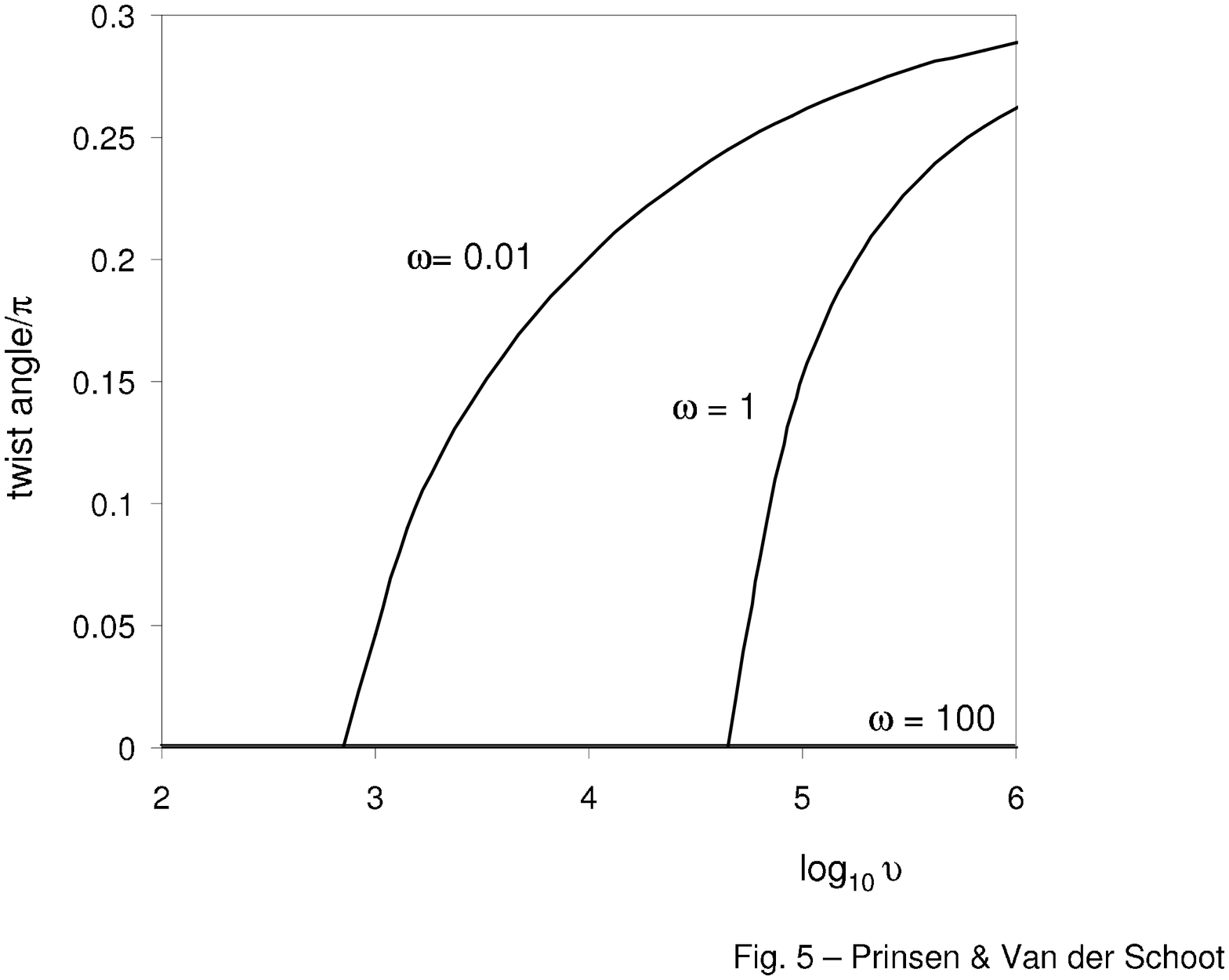, angle=-90, totalheight=11 cm}
\end{center}
\end{figure}

\begin{figure}
\begin{center}
\leavevmode \epsfig{file=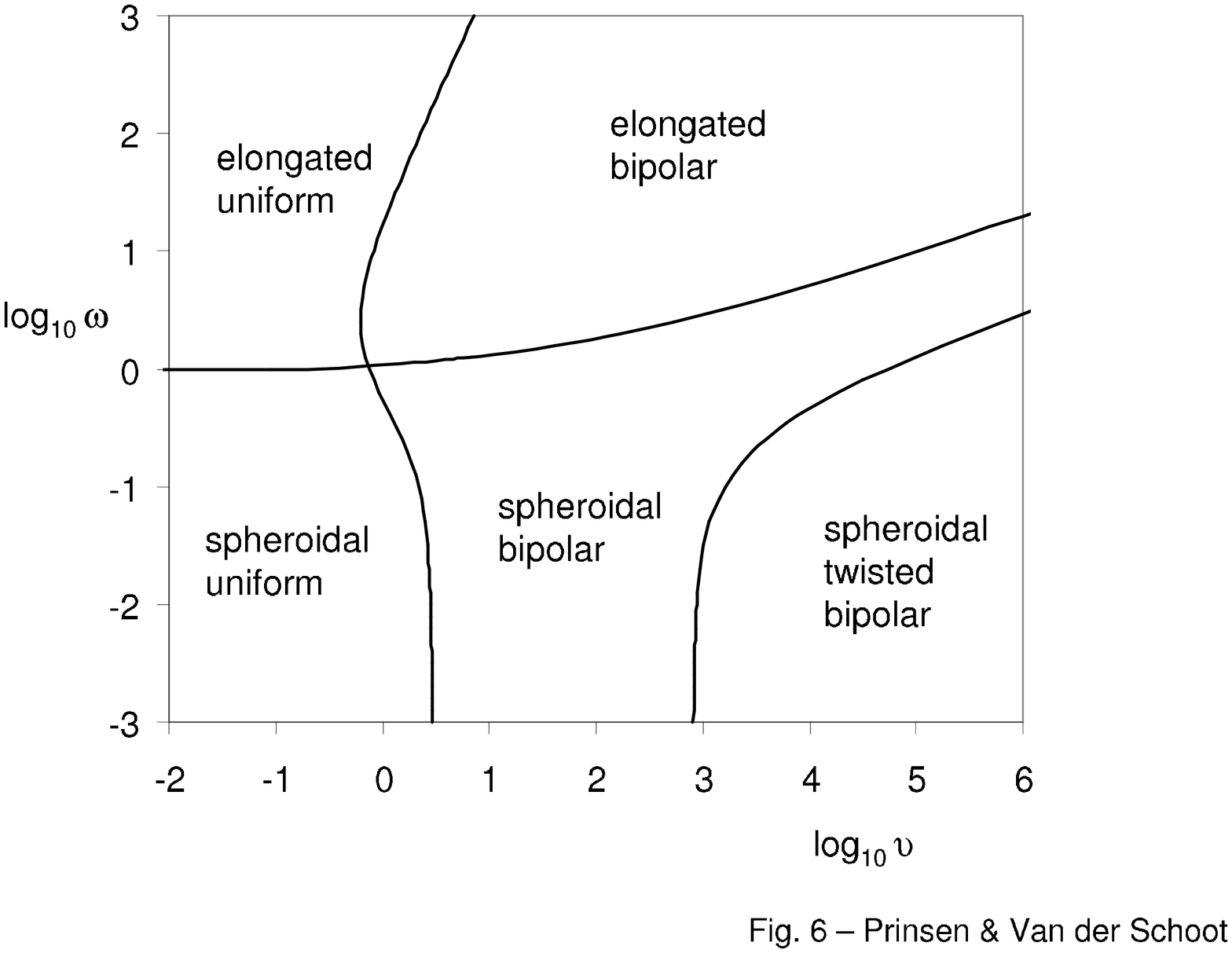, angle=-90, totalheight=11 cm}
\end{center}
\end{figure}

\begin{figure}
\begin{center}
\leavevmode \epsfig{file=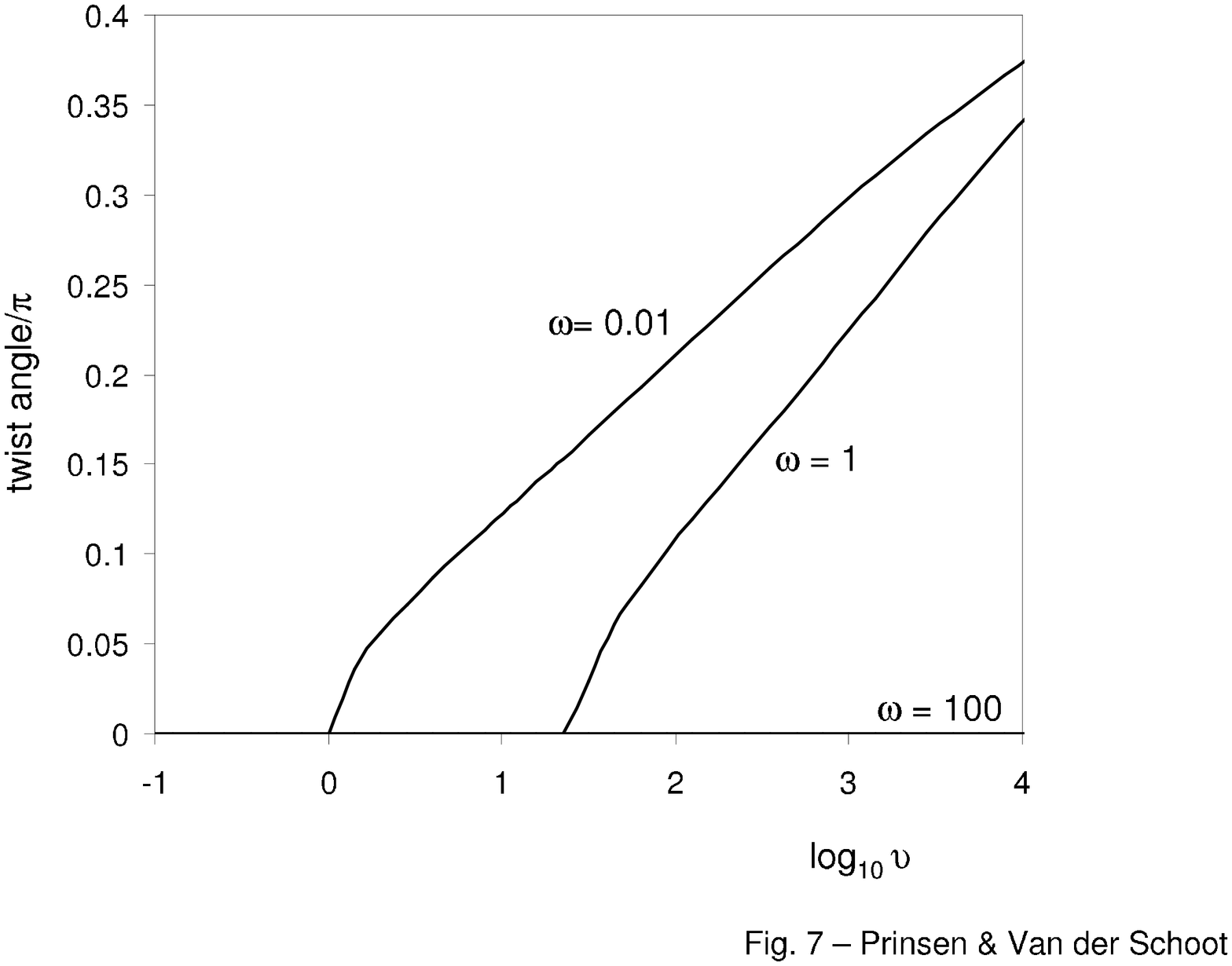, angle=-90, totalheight=11 cm}
\end{center}
\end{figure}

\begin{figure}
\begin{center}
\leavevmode \epsfig{file=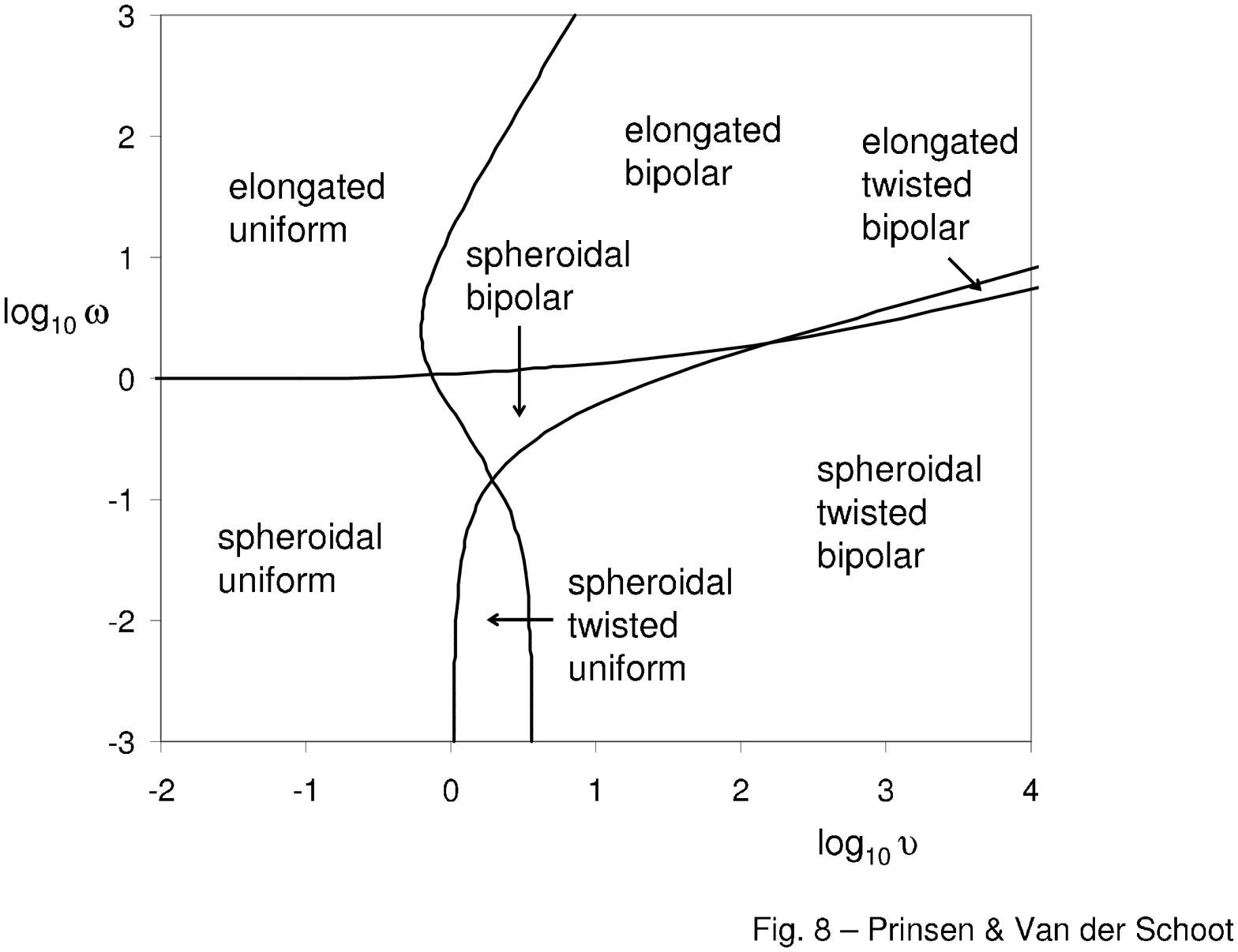, angle=-90, totalheight=11 cm}
\end{center}
\end{figure}

\begin{figure}
\begin{center}
\leavevmode \epsfig{file=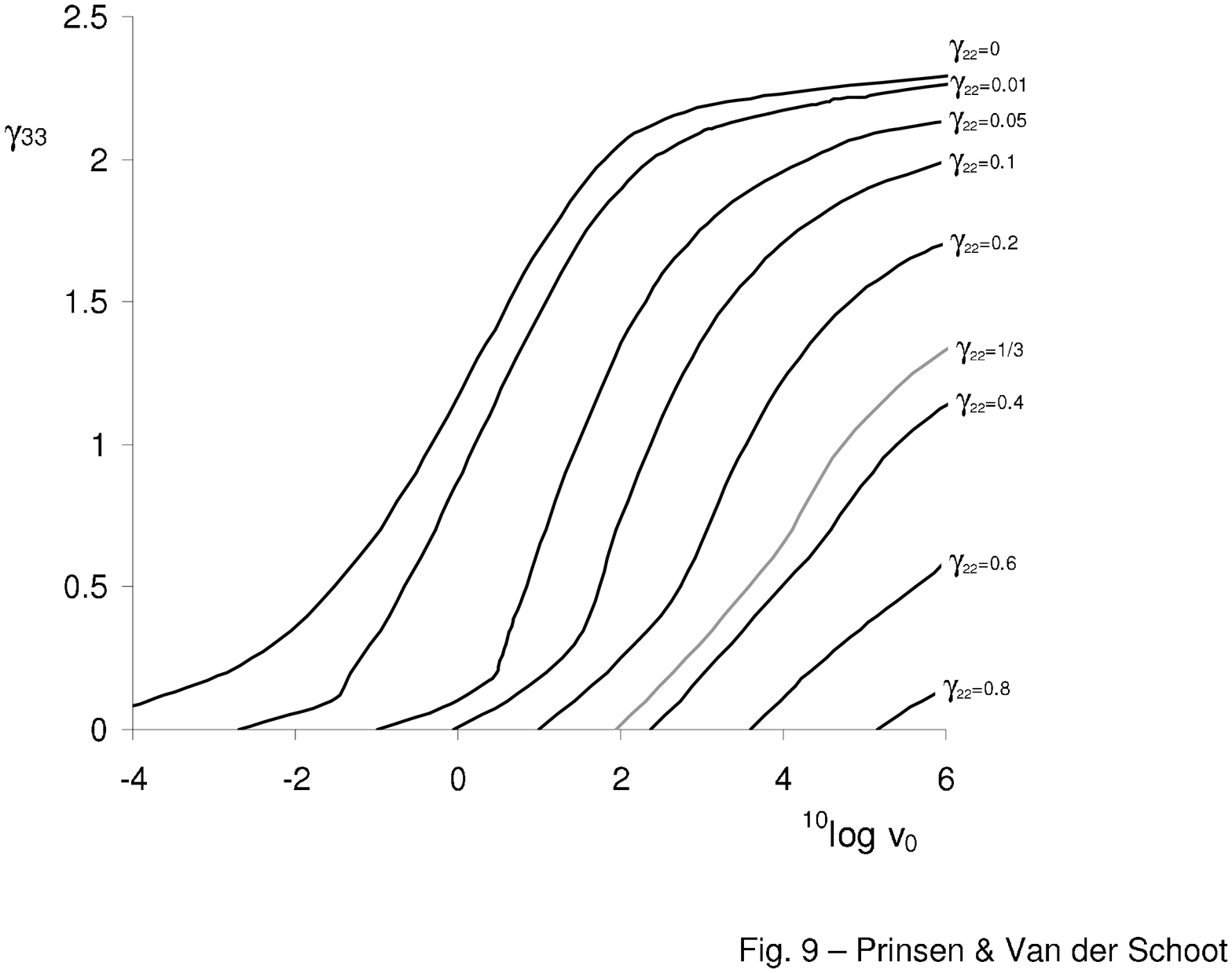, angle=-90, totalheight=11 cm}
\end{center}
\end{figure}

\end{document}